\RequirePackage{amsmath}

\documentclass[conference]{IEEEtran}

\usepackage{mdframed}
\usepackage{xcolor}
\usepackage{cite}
\usepackage{url}
\usepackage[utf8]{inputenc}
\usepackage{paralist}
\usepackage{float}
\usepackage{graphicx}
\usepackage{hyperref}

\usepackage{framed}
\usepackage{amsthm}
\usepackage{amssymb}

\theoremstyle{definition}

\newcommand{\noteTaeho}[1]{{$\langle${\textcolor{blue}{\textbf{#1}}}$\rangle$}}

\makeatletter
\def\blfootnote{\gdef\@thefnmark{}\@footnotetext}
\makeatother

\begin{document}
\title{Energy-recycling Blockchain with Proof-of-Deep-Learning}

\author{\IEEEauthorblockN{Changhao Chenli$^*$, Boyang Li$^*$, Yiyu Shi, Taeho Jung}
\IEEEauthorblockA{\textit{Department of Computer Science and Engineering} \\
\textit{University of Notre Dame}\\
Notre Dame, Indiana, USA \\
\{cchenli,Boyang.Li.258,yshi4,tjung\}@nd.edu}
}

\iffalse
\author{Changhao Chenli$^*$, Boyang Li$^*$, Yiyu Shi, Taeho Jung}
\institute{University of Notre Dame, Notre Dame IN 46556, USA \\
\email{\{cchenli,Boyang.Li.258,yshi4,tjung\}@nd.edu}
}
\fi

\maketitle

\blfootnote{* Authors with equal contributions listed alphabetically with last names.}

\iffalse
\noteTaeho{Double-click at the previewd pdf will bring you to the corresponding source.}

\noteTaeho{Do not care about the page limit. Just include all necessary and relevant contents.}
\fi

\begin{abstract}

%Blockchain, as a revolutionary technology of distributed and tamper-proof public ledger, has been actively studied in both academia and industry owing to its huge potential value. One of the most important mechanisms in the blockchain is its Proof of Work (PoW) mechanism, which is an essential step for generating and validating new blocks and realizing the DAO. However, a
An enormous amount of energy is wasted in Proof-of-Work (PoW) mechanisms adopted by popular blockchain applications (\textit{e.g.,} PoW-based cryptocurrencies), because miners must conduct a large amount of computation. %Such a PoW inherently drains a huge amount of energy and generate considerably much CO2 accordingly. 
Owing to this, one serious rising concern is that the energy waste not only dilutes the value of the blockchain but also hinders its further application.
In this paper, we propose a novel blockchain design that fully recycles the energy required for facilitating and maintaining it, which is re-invested to the computation of deep learning. We realize this by proposing Proof-of-Deep-Learning (PoDL)  such that a valid proof for a new block can be generated if and only if a proper deep learning model is produced. We present a proof-of-concept design of PoDL that is compatible with the majority of the cryptocurrencies that are based on hash-based PoW mechanisms. Our benchmark and simulation results show that the proposed design is feasible for various popular cryptocurrencies such as Bitcoin, Bitcoin Cash, and Litecoin.

%\noteTaeho{Need to change the claim}

%s such as deep learning. With this energy recycling, more blockchain-based applications will be enabled without wasting energy, and a huge amount of energy needed for the blockchain will become available to the AI community for conducting complex computation task.
%To fulfill this goal, we ask and study four research questions:

\begin{IEEEkeywords}
Blockchain, Sustainability, Deep Learning
\end{IEEEkeywords}
%\keywords{}
\end{abstract}

\section{Introduction}

%\subsubsection*{Energy waste in blockchain}

In the past decade, blockchain technology has been successfully applied in different fields and potentially will be applied in more critical areas. However, the current mechanism consumes a huge amount of energy for conducting the computation needed for maintaining the security guarantee of the system, and various concerns are arising accordingly. ``The cryptocurrency uses as much CO$_2$ a year as 1 million transatlantic flights. We need to take it seriously as a climate threat'' according to the Guardian \cite{hern_2018}. Because of such concerns, ``Bitcoin's need for electricity is its Achilles Heel'' according to the Forbes \cite{coppola_2018}. 
%With rapidly increasing hash difficulty and numbers of transactions, the energy consumption of Bitcoin's blockchain networks has been increasing rapidly and steadily.
\iffalse

\begin{figure}[t]
    \centering
    \includegraphics[width=.8\linewidth]{chart.pdf}\vspace{-10pt}
    \caption{Energy consumption of Bitcoin from Fabruary 2017 to October 2018.}
    \label{fig:energy}
\end{figure}
\fi
According to \cite{digiconomist}, the energy consumption of Bitcoin has been steadily increasing until late November 2018 when the Bitcoin price dropped suddenly. Even after the drop, more than 50 TWh (trillion watt hour) per year is consumed just for maintaining the blockchain underlying Bitcoin.  This amount is from Bitcoin only, and the total energy consumption by all applications based on blockchain will be much more than that. 

The main issue is that all energy is wasted to some extent. The majority of the energy is being consumed by the hash calculation in PoW-based blockchains. Proof-of-Capacity (PoC) is proposed to address this issue (\textit{e.g.,} Burstcoin \cite{burstcoin}), however it does not completely solve the problem because storage resources are wasted instead.

Primecoin \cite{king2013primecoin} uses prime number finding instead of hash calculation as PoW, and miners seek special sequences of prime numbers (Cunningham chains). However, the application of those numbers  is limited in cryptographic protocols. Gridcoin \cite{gridcoin-white-paper}, Golem \cite{home-golem}, and FoldingCoin \cite{Foldingcoin} are cryptocurrencies that distribute rewards to miners based on the amount of scientific computation they performed. Though being similar, the Proof-of-Deep-Learning (PoDL) proposed in this paper differ from them substantially. Miners' computing ability does not make blockchain secure in those systems, however PoDL is an improved PoW-like consensus mechanism with which deep learning power of honest miners provide tamper-proofness. Owing to this, PoDL can be deployed in any PoW-based blockchain applications, recycling their miners' energy for deep learning. 
\iffalse

, and Gridcoin \cite{gridcoin-white-paper}, where performing scientific computation is part of block mining, are relevant to the Proof-of-Deep-Learning (PoDL) proposed in this paper. % the only cryptocurrencies that encourage miners to conduct useful computation tasks.
 Miners in Gridcoin perform scientific computation, and their rewards are proportional to the computation amount. Though being similar, Gridcoin's goal is orthogonal to this paper. Block rewards in Gridcoin is proportional to the amount of performed computation, 

while our goal is to present an alternative PoW mechanism for recycling energy consumed by existing blockchain applications.
\fi
Proof-of-Stake  (\textit{e.g.,} Nxt \cite{nxt}) or Proof-of-Important (\textit{e.g.,} NEM \cite{nem}) are alternative consensus mechanisms with less energy consumption. However, their principle is orthogonal to that of PoW, and the `block mining' does not involve computation. Therefore, these are orthogonal to our work.

%and Proof-of-Capacity (PoC) for Burstcoin \cite{burstcoin}. PoS mechanisms (or other variants such as Proof-of-Importance \cite{nem}) is an orthogonal mechanism with different principle: decides new blocks based on stakes/reputation, 

%Proof-of-Stake as the consensus mechanism, and it is orthogonal to our effort of making PoW-based blockchain more sustainable by recycling the consumed energy.

%are not useful for cryptographic protocols because 

%large prime number finding substitutes hash calculation in PoW, however the prime numbers found in Primecoin are not useful in cryptographic protocols. In Gridcoin, block rewards miners receive are proportional to the amount of scientific 

%. Miners in Primecoin generate a prime number array called Cunningham chain, however, the prime numbers found in Primecoin are not useful in cryptographic protocols.

We present a novel design of blockchain which reinvests the energy consumed by blockchain maintenance in computation tasks of deep learning. This is done by introducing Proof-of-Deep-Learning (PoDL) mechanism which forces miners to perform deep learning training and present trained models as proofs.
The contributions of this paper are summarized as follows. % Among existing consensus mechanisms for blockchain (PoW for Bitcoin \cite{nakamoto2008bitcoin}, PoS for Nxt \cite{nxt}, PoC for Burstcoin \cite{burstcoin} \textit{etc.}), PoDL is the first one that encourages miners to perform tasks useful for other applications.
%Our PoDL mechanism resolves the issues of ASIC devices as well.  ASIC is traditionally considered adverse to the blockchain ecosystem. It will in fact be beneficial in our blockchain because the ASIC towards our newly designed PoW mechanism will be aimed for deep learning training, and it will contribute to the development of the better hardware.% will be beneficial both to blockchain-based systems and to the ML field. In blockchain applications, by using advanced ML-aimed ASIC, the block interval will decrease and get close to the real interval, which makes our system more capable and helps all blockchain-based systems get rid of useless ASIC machines.  As for ML field, it is a powerful tool for getting the result in a rapid speed.
\iffalse
Our energy recycle project will be helpful in this trend and both of AI and blockchain society will take benefit from this novel mechanism. The concept of Primecoin proved that it is unnecessary to run hash algorithm as proof of work, but their mechanism can hardly adopt to more helpful algorithm. 

Our task is to recycle the ‘wasted’ energy and the blockchain ecosystem will still behave the same. The methodology will be introduced in the next section. Our concept is to share the calculation power to other purpose such as machine learning. 
\fi

(1) We present the first consensus mechanism, PoDL, that maintains blockchain via deep learning instead of useless hash calculation; %Majority of computing resources are invested to  training proper deep learning models.
(2) Our PoDL can be applied to any cryptocurrency based on PoW mechanisms because we only incrementally add components to block headers;  %Additionally, it can be applied to any blockchain-based applications as well.
(3) Our experiment shows that the design is feasible for cryptocurrencies whose block intervals are much greater than 10 seconds.

\section{Preliminaries}
\subsection{Proof of work (PoW) and Block Mining}

%\noteTaeho{Brief introduction to PoW and block mining for people without knowledge.}

\textit{Proof of work (PoW)} \cite{dwork1992pricing} is used commonly in many cryptocurrencies (Bitcoin, ZCash, Monero, Litecoin \textit{etc.}), where miners need to find a hash value smaller than some threshold, and this involves brute-force search over a large search space. Therefore, plenty of computation resource is needed in the PoW mechanism of those cryptocurrencies.
\textit{Block mining} is the process of creating a block with a valid hash value (\textit{i.e.,} less than a small threshold). Miners who create blocks with valid hash values are rewarded for their work. More specifically, they are allowed to insert one Coinbase transaction which creates and sends certain amount of \textit{block reward} to any address specified by the miner. %Therefore, this process is called \textit{block mining}.

\subsection{Deep Learning (DL) and its Training}

Deep learning \cite{lecun2015deep} (DL) outperforms traditional machine learning algorithms dramatically in many areas. To achieve a proper DL model, one needs to provide a dataset (called \textit{training dataset}) and train the model. The training is composed of two algorithms, \textit{feed-forwarding} and \textit{back-propagation}, that are interchangeably executed. %In the former algorithm, neural network's outputs are calculated given an input record (\textit{i.e.,} features with a label), and the error is back-propagated in the latter algorithm to calibrate neural network parameters. 
We say an \textit{epoch} is finished when a pair of feed-forwarding and back-propagation are interchangeably performed through the  neural network exactly once for every record in the training dataset. Multiple epochs are repeated in training, and the accuracy of a trained model can be tested with another dataset without overlaps (called \textit{test datset}). Such a training is an approximation algorithm based on hill climbing that seeks local optima in the entire model parameter space. Efficient algorithms that find global optimum are unknown yet, and this is the core of our PoDL.

\section{Our Energy-recycling Blockchain with PoDL}

%In this section we are going to introduce our solution to the current problem. 

%Current block mining consumes a large amount of computing resources on calculating meaningless hash values. 
We propose to recycle the energy consumed in the block mining by introducing Proof-of-Deep-Learning (PoDL) with DL training. Namely, we let miners train DL models, and the block generated by the miner who trained a \textit{proper DL model} will be accepted by full nodes. In the block chain with PoDL, we have an extra stakeholder besides miners and full nodes:  \textit{model requester} who outsources DL model training to miners. The goal of this paper is to present a proof-of-concept design, and we consider the simplest model where there is only one model requester who provides training/test datasets that describe the desired model. Because the model requester's goal is to get the best DL model, we assume s/he will be a semi-honest adversary who does not collude with anyone.

\subsection{Overview of New Blockchain with PoDL}

\noindent \textit{$\bullet$ Block acceptance policy.} When miners submit blocks and block headers, we let them submit trained DL models. Then, we let full nodes choose the block that is valid and also comes with the model with the highest accuracy when validated with the test datasets. Full nodes are asked to validate models on test datasets on their own to calculate the accuracy. In order to prevent Denial-of-Service attack, we let miners self-validate their models and report their models' accuracy as well. Full nodes are asked to start their validation from the models with the highest accuracy first and stop when they find the first model whose validated accuracy is same as the claimed one. This replaces the PoW validation, and we do not require the hash of the block header to be smaller than threshold values. To tie blocks and DL models, we require that models be hashed and that hash values be included in block headers.

\noindent \textit{$\bullet$ Preventing model overfitting.} If test datasets are available to miners, they are motivated to cheat by training DL models directly on test datasets, \textit{i.e.,} overfitting the model. To prevent this, we set up two phases between blocks. In the first phase, the model requester releases training datasets to miners for their training, and they do not release test datasets until the second phase. Miners will become able to validate their accuracy and submit the models to full nodes after the first phase is over. The following mechanism will prevent miners from continuing the training in the second phase.

%miners to validate the accuracy of their trained models and then to compete with others by submitting their trained models and the blocks.

\noindent \textit{$\bullet$ Preventing model stealing and training in the second phase.} Miners may cheat by (1) stealing others' DL models published in the second phase or by (2) further training DL models with the released test datasets for higher accuracy (\textit{i.e.,} model overfitting). To prevent these, we require that miners release the block headers in the first phase if they want to compete in the second phase, and the headers serve as the \textit{commitment} of their models. In the second phase, full nodes will validate the blocks and models whose block headers have been submitted in the first phase only. By doing so, if miners steal others' models or retrain their models, their new models' hash value will be different, and their block headers will be different from what full nodes received in the first phase.

\noindent \textit{$\bullet$ Blockchain verification:} To verify the whole blockchain (\textit{e.g.,} when Initial Block Download occurs), full nodes must ensure the accepted DL models are trained from training datasets only and their accuracy in the test datasets is same as the claimed one. To provide such verifiability, miners are asked to submit the parameters necessary for repeating the training: hyperparameter, initial weights, number of epochs \textit{etc.}. With these, full nodes are able to repeat the training to determine whether the accepted model can be reconstructed from the training datasets only. Furthermore, they can verify the claimed accuracy with test datasets. 

%besides the blocks and headers, our blockchain also keeps DL models and datasets. Anyone who wishes to verify the whole blockchain can verify that the accepted DL model by validating its accuracy in the test dataset.

\iffalse
Adversaries may try to double spend by submitting a number of blocks/headers/models at the same time, arguing that his/her branch is created by the network delay. 

have access to test datasets since they launch the attack for a set of blocks $\{b_t,b_{t+1},\cdots\}$ after they are confirmed, \textit{i.e.,} after test datasets have been released. To ensure submitted models are trained from training datasets only, we require that miners submit the parameters necessary for repeating the training: hyperparameter, initial weights, number of epochs \textit{etc.}. With these, full nodes are able to

Because training algorithms seek local optima with certain randomness, it is challenging to increase the accuracy beyond certain point. If only the highest-accuracy models are accepted, it is challenging to further improve the accuracy beyond it. If adversaries wish to double spend in our blockchain, they need to present DL models that has a higher accuracy than the best one among all other miners' models for multiple blocks. We conjecture that this is extremely challenging unless the adversary possesses more than 51\% of the computing resources for DL training. 
\fi

\subsection{Blockchain Description with PoDL}

\noindent \textbf{Phase 1 for determining block $b_t$ at height $t$:} Given training datasets released by the model requester, miners train DL models \textit{without knowing test datasets} as part of PoDL. The miners generate the block and the header by following the rule of the underlying blockchain system (\textit{e.g.,} generate transactions and merkle trees of blocks in cryptocurrencies) and including the hashed model, and they submit the block header to full nodes by the end of Phase 1. %Examples of a block header and a blockchain are shown in Fig. \ref{fig:example-header} and Fig. \ref{fig:example-chain}.

\noindent \textbf{Phase 2 for determining block $b_{t}$ at height $t$:} The model requester releases test datasets, and miners validate their trained models and submit the highest-accuracy ones to full nodes along with the block and the block header. Then, full nodes choose and validate the accuracy of submitted DL models in the decreasing order of the accuracy claimed by the miners, and accept the first one (as well as the corresponding block and header) that has the claimed accuracy. In the case of a tie, full nodes follow the policy of underlying blockchain system (\textit{e.g.,} accept the one which arrived earlier as in Bitcoin). Full nodes ignore all models whose block headers are not received in the first phase, and furthermore, full nodes do not accept other blocks at height $t$ once Phase 2 for $b_t$ is finished.

\iffalse

\begin{enumerate}
    \item
    Training and Submission Phase: From the miners' eyesight, instead of calculating  difficult hash values, they will start training a DL model after the previous block was added onto the blockchain and the training data set was published by full nodes. After some time, miners will submit the hash value of their models and the full nodes will maintain a log for collecting these submissions.
    \item Validation and Generation Phase: After the submissions were collected by the full nodes, the testing data set will be published by the full nodes. The miners will then start to varify the accuracy of their models and publish their model as well as their block (including previous block's information, his own model and some transactions), his model and related accuracy. The full nodes will firstly check whether the models had been submitted during the first phase and then verify the model with highest accuracy among all the submissions. If the verification is correct, then the new block will be added onto the blockchain.

\end{enumerate}
\fi

\iffalse
\begin{figure}[t]
    \centering
    \includegraphics[width=.8\linewidth]{chart.pdf}
    \caption{An example block header with our PoW mechanism.}
    \label{fig:example-header}
\end{figure}

\fi

\begin{figure}[t]
    \centering
    \includegraphics[width=\linewidth]{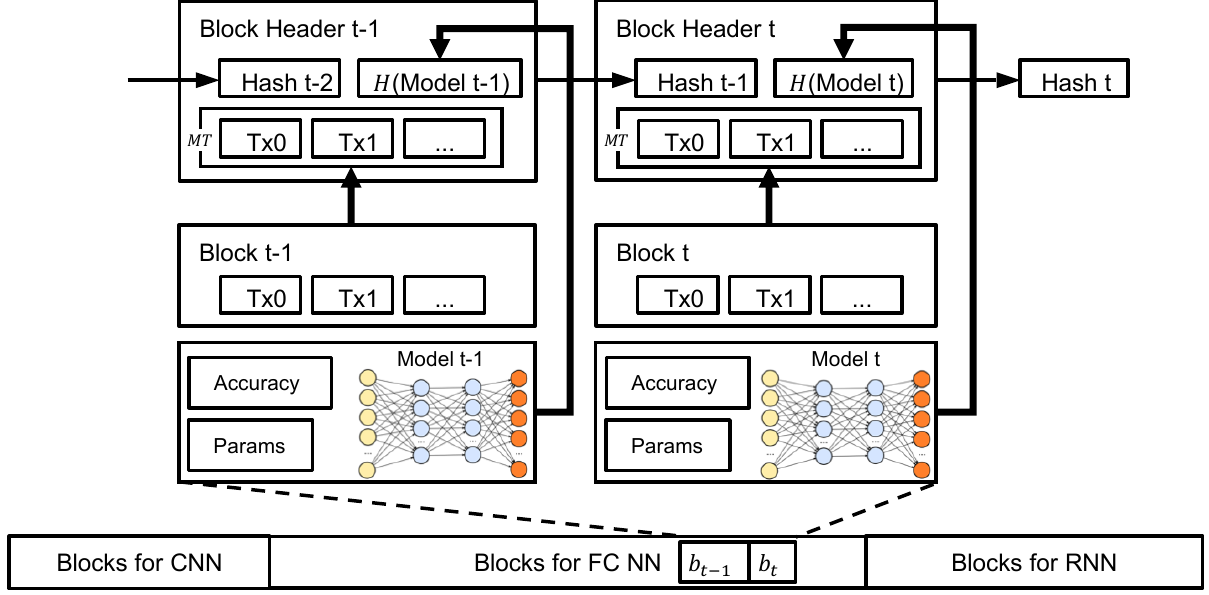}\vspace{-5pt}
    \caption{A toy example that trains CNN, Fully Connected NN, and RNN.}
    \label{fig:example-chain}\vspace{-10pt}
\end{figure}

\noindent \textbf{Dealing with short training time:}
Block generation rates are controlled to be constant on average in many cryptocurrency systems (\textit{e.g.,} 10 minutes in Bitcoin and 2.5 minutes in Litecoin). Therefore, miners train for a short period time only in Phase 1, yielding low accuracy increment. We present two mechanisms to address this. First, the model requester does not collect the model until the model accuracy does not increase significantly after multiple blocks, and s/he releases new test datasets for different blocks. By doing so, training for one DL model spans across multiple blocks, and a good-enough model is achieved at the end. Besides, Phase 2 for $b_t$ and Phase 1 for $b_{t+1}$ may happen concurrently. Namely, after validating the accuracy of trained model for $b_t$ (which is Phase 2 for $b_t$), miners immediately start training for $b_{t+1}$ (which is Phase 1 for $b_{t+1}$). This is acceptable as long as the model requester provides fresh test datasets for every block. %, and this allows miners to constantly train DL models.% By doing so, Phase 2 for $b_{t}$ and Phase 1 for $b_{t+1}$ occurs concurrently for any $t$, and we have optimum time utilization since miners constantly train DL models.
Note that one training dataset is reused across multiple blocks, therefore the miners need to access the training dataset only once per model.

\iffalse

\noindent \textbf{Overlapping phases:}
 Phase 2 for a block $b_t$ and Phase 1 for the next block $b_{t+1}$ are independent from each other since they deal with different models, therefore they may happen concurrently. After validating accuracy of trained models for $b_t$ and submitting  models/blocks/headers to full nodes (which is Phase 2 for $b_t$), miners start training new models for $b_{t+1}$ immediately (which is Phase 1 for the next block $b_{t+1}$). By doing so, Phase 2 for $b_{t}$ and Phase 1 for $b_{t+1}$ occurs concurrently for any $t$, and we have optimum time utilization since miners constantly train DL models after validating the accuracy (whose overhead is negligible compared to training).
 
\fi

The rest of the blockchain remains the same. Finally, we present an example blockchain with our new PoW mechanism in Fig. \ref{fig:example-chain}, where $MT$ stands for the root of the Merkle Tree.

\subsection{Properties of Our Design with PoDL}

\noindent \textbf{Block reversibility:} Because of our block acceptance policy, accepted DL models have lower accuracy in earlier blocks, and the DL models have higher accuracy in later blocks. Accordingly, it becomes much more challenging to present a model whose accuracy is higher than the accepted ones. Therefore, the previous blocks become hardly reversible only after the models with high-enough accuracy appear in the blockchain. Due to this, whether blocks are reversible does not depend on the number of confirmations. Rather, it depends on the highest accuracy of the model along the blocks.% Users are encouraged to treat blocks as confirmed if the accepted models' accuracy is high enough.
\\\textbf{Hardness of double spending:}
Firstly, full nodes in our blockchain accept the blocks in Phase 2 if and only if their headers are received in Phase 1. Therefore, even if adversaries have access to test datasets after a block $b_t$ is confirmed, they are unable to submit new blocks with new models because the corresponding block header does not exist in the list of Phase 1 for $b_t$ as long as full nodes are honest.

Even if the majority of the full nodes collude with miners, double spending without 51\% computing resources is still a low-probability event. Training algorithms seek local optima with certain randomness because no known algorithms can find the global optimum. Therefore, if only the highest-accuracy models are accepted, it is challenging to further improve the accuracy beyond it (also see Fig. \ref{fig:accuracy-per-block} in our simulation). If adversaries wish to double spend in our blockchain by controlling majority of the full nodes, they must present a longer sequence of blocks where all blocks must contain DL models with higher accuracy. Furthermore, the DL models' training must be repeatable with training datasets only as well.  Owing to the randomness of the training performance that depends on the random choice of hyperparameters and initial weights, we conjecture that this is extremely challenging unless the adversary possesses more than 51\% of the computing resources for DL training. Adversaries having good hyperparameters may have an advantage for reversing the blocks, however those parameters will be published to other miners as well, making it hard to reverse blocks again.
\\\textbf{Datasets provision:} Training and test datasets for DL may have large volumes, however these are necessary for blockchain verification. The storage burden will be prohibitively high if datasets are stored in the blockchain, therefore we assume model requester will provision datasets properly (\textit{i.e.,} by following the release time for different blocks). Model requesters are motivated to play this role as their goal is to get the best model.
\\\textbf{Storage burden:} DL models' sizes vary from 100KB to 10GB, and storing all model parameters including those for training repeatability can be a huge burden. However, various techniques can be used to reduce the sizes without affecting the accuracy too much \cite{han2015deep,gong2014compressing}, and we may limit the model sizes to a common one, \textit{e.g.,} 10MB/model in \cite{han2015deep,gong2014compressing}. Furthermore, because the tamper-proofness is guaranteed by high-accuracy models only, we can free the storage by removing models with low accuracy. %In this case Considering that the model requester will accept the trained model when the accuracy does not increase significantly with new blocks, we only need to keep as many models as we need for preventing double spending (\textit{e.g.,} 5-10 blocks). 
The later blocks with high-accuracy models will still prevent the double spending.
%\noteTaeho{Need to describe that training/test datasets will be stored at model requester's side. Need to justify why model requester needs to and will be motivated to provide training/test datasets.}
%in both storage and network traffic, which is not negligible. However, considering the steady improvement in storage/network capacity, it will become manageable in the near future.
%\noteTaeho{Also need to describe the network delay of training dataset propagation along the network. It's acceptable if we keep using the same training datasets over and over.}
\\\textbf{Network delay:} %Although we have introduced model requester into our PoDL mechanism to make training datasets and test datasets available all the time, we still need to consider the impact brought by doing this. 
Blocks submitted to full nodes include DL models and training parameters, therefore full nodes will experience extra network delay. Besides, miners experience extra delay as well owing to the retrieval of training/test datasets. However, note that the same training datasets are used for multiple blocks, without needing retrieval at every block. New test datasets need to be retrieved in every Phase 2, however the phase overlaps with Phase 1 for the next block, and it does not affect the block generation rate as long as the network delay is smaller than the block interval. Besides, the block acceptance is determined by the accuracy of DL models rather than their arrival time (except for tied models), the impact of test datasets' network delay is minimal.
\iffalse
Additionally, training/test datasets need to be propagated

, and the time difference could not be neglected in sometime, which therefore caused many problems such as forking. In our mechanism, as the test datasets are used for verifying, which is the duty of full nodes', we only need to make miners train on the same training datasets again and again, and this idea is also common in DL area as most models are kept training on the same training dataset until it reach certain accuracy or time limit. This setting is also correlated to the validation process, where each new block will contain not only the model but also some related parameters, including how to get the model and so on. In that way, as everyone has the same training dataset, who is going to verify a certain block together with the model it contains will be very easy to implement, and can also reduce the memory cost of model requester.
\fi
\\\textbf{Impact to ASIC devices:}
%Our PoDL mechanism resolves the issues of ASIC devices. 
ASIC is traditionally considered adverse to the blockchain ecosystem, but it will in fact be beneficial in the blockchain with PoDL because ASIC devices will be designed for deep learning training, and it will contribute to the development of better hardware.

\section{Feasibility Validation by Experiments}

\subsection{Experiment Setting}

The experiment for block generation and validation was conducted with a laptop with Intel i7-6700. We implemented blockchain functions (\textit{e.g.,} transaction/block generation, hash calculation) based on \cite{dvf2017} under Python 3.6.
The deep learning experiment was conducted with a desktop with i7-6850K, 24GB RAM, and two GTX 1080Ti GPUs. A single-word command recognition model was trained with TensorFlow using a dataset including 105,000 audio samples \cite{warden2018speech}. 

%\noteTaeho{Need Boyang's experiment description}

% dataset: 
% https://storage.cloud.google.com/download.tensorflow.org/data/speech_commands_v0.02.tar.gz

% if we want YOLO, this is the citation. 
%
% @article{yolov3,
%   title={YOLOv3: An Incremental Improvement},
%   author={Redmon, Joseph and Farhadi, Ali},
%   journal = {arXiv},
%   year={2018}
% }

\subsection{Benchmark Tests}

Instead of constantly calculating hash values, miners will constantly conduct DL training in our blockchain. Therefore, DL training itself is the counterpart of hash brute force rather than extra overhead compared to existing blockchain systems. The extra computation tasks brought by our new PoDL mechanism are: (1) Miners' hash calculation for $H$(model) at Phase 1; 
(2) Full nodes' sorting by accuracy at Phase 2.;
(3) Full nodes' validation for $H$(model)'s correctness and search for the block header at Phase 2.;
(4) Miners' accuracy determination at Phase 2.;
(5) Full nodes' accuracy verification at Phase 2.;
(6) Full nodes' full verification at Initial Block Download.
Owing to lack of sufficient data, we omit the evaluation for (6) which requires a large number of models as well as their training parameters. Since (6) is a one-time process, its impact is much smaller than the rest.
For the rest,
 we measure the elapsed time for conducting those tasks and compare it with common block intervals, which explains how much of miners' time is devoted to DL training, \textit{i.e.,} effectiveness of our energy recycling.

\begin{table}[t]
    \centering
    \caption{Benchmark of Hash, Hash table, and Sorting}\vspace{-5pt}
    \label{tab:negligible}
    \begin{tabular}{c|c|c}
    \hline\hline
        \textbf{Hash} & \textbf{Hash table} & \textbf{Sorting} \\
        (SHA-256)& (Google Dense Map) & (Quicksort)
        \\\hline
        5.9 ms/MB \cite{hash-benchmark} & 89.75ms/1M inserts & 154.9 ms/1M objects \cite{sorting-benchmark}  \\
        & 16.25ms/1M reads \cite{hashtable-benchmark} & \\
        \hline\hline
    \end{tabular}\vspace{-10pt}
\end{table}

\iffalse
For the comparison on direct overhead caused by our design, we performed our experiments on our modified blockchain systems. We ran our experiments for 1000 times to get the average value. Fig. \ref{fig:exp1} shows the block validation time cost based on our simulation system. We can see from the chart that the overhead brought by our design is about 1.9 seconds. Among 1000 tests, the average time cost of block validation was about 1.96002132058144s, which is acceptable to many cryptocurrency systems.% compared to the time limit of block generation interval (10 minutes).  
\fi

Hash calculation is involved in (1); sorting is needed in (2); sorting and searching is needed in (3). For those, we present existing benchmark results in Table \ref{tab:negligible} to show their overhead. Note the load factor for the hash table in \cite{hashtable-benchmark} is $\alpha=0.38$.  Generally speaking, their extra overhead is negligible compared to block intervals of any cryptocurrency. % when block intervals are much larger than 1 second. Therefore, PoDL will work well
%\noteTaeho{average over at least 10 repetitions}

\begin{figure}[t]
    \centering 
    \includegraphics[width=1\linewidth]{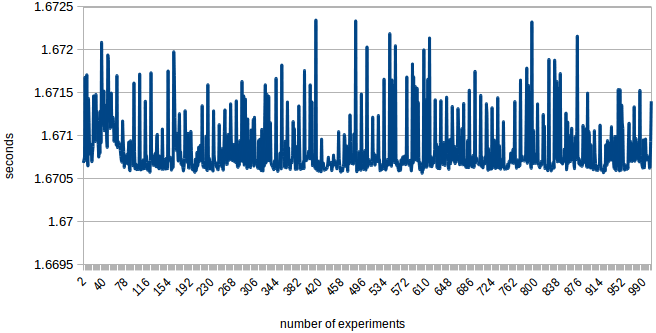}\vspace{-10pt}
    \caption{Model validation time among 1000 models}
    \label{fig:exp1}\vspace{-10pt}
\end{figure}

(4) and (5) involve feed-forwarding on DL models. Miners need to run feed-forwarding algorithm for once for every record in the test datasets. Full nodes need to run that algorithm for as many times for every record as the number of models they need to validate in Phase 2, but this will be small because full nodes will stop validating the models as soon as they find the model with the claimed accuracy. We present the elapsed time for model validation in Fig. \ref{fig:exp1}. The average over 1,000 repeated block validations is 1.96 seconds, which is negligible when compared to block intervals of some popular cryptocurrencies (\textit{e.g.,} 10 minutes for Bitcoin and Bitcoin Cash, 2.5 minutes for Litecoin), meaning that PoDL works well with those cryptocurrencies because most energy can be recycled. If block intervals are smaller (\textit{e.g.,} 10-19 seconds for Ethereum), PoDL will be less effective.

\iffalse
compared to block intervals of some popular cryptocurrencies (10 minutes for Bitcoin and Bitcoin Cash, 1 minute for Monero and 2.5 minutes for Litecoin), however the overhead is not negligible for some other popular cryptocurrencies (10-19 seconds for Ethereum and 3.5 seconds for Ripple). 
\fi

\iffalse

\begin{figure}[t]
    \centering 
    \includegraphics[width=0.4\textwidth]{exp_3_1.png}\vspace{-10pt}
    \caption{Comparison of different models}
    \label{fig:exp3}
\end{figure}

For the last experiment, we chose 3 models in different sizes, 1Mb, 3.5Mb and 5Mb, and performed the same experiments for 1000 times same to the first one. From Fig. \ref{fig:exp3}, we can see that the difference is really small as hash calculation speed varied slightly on different size of the input. 

\fi

\subsection{Simulation for Accuracy Growth}

\begin{figure}[t]
    \centering 
    \includegraphics[width=0.8\linewidth]{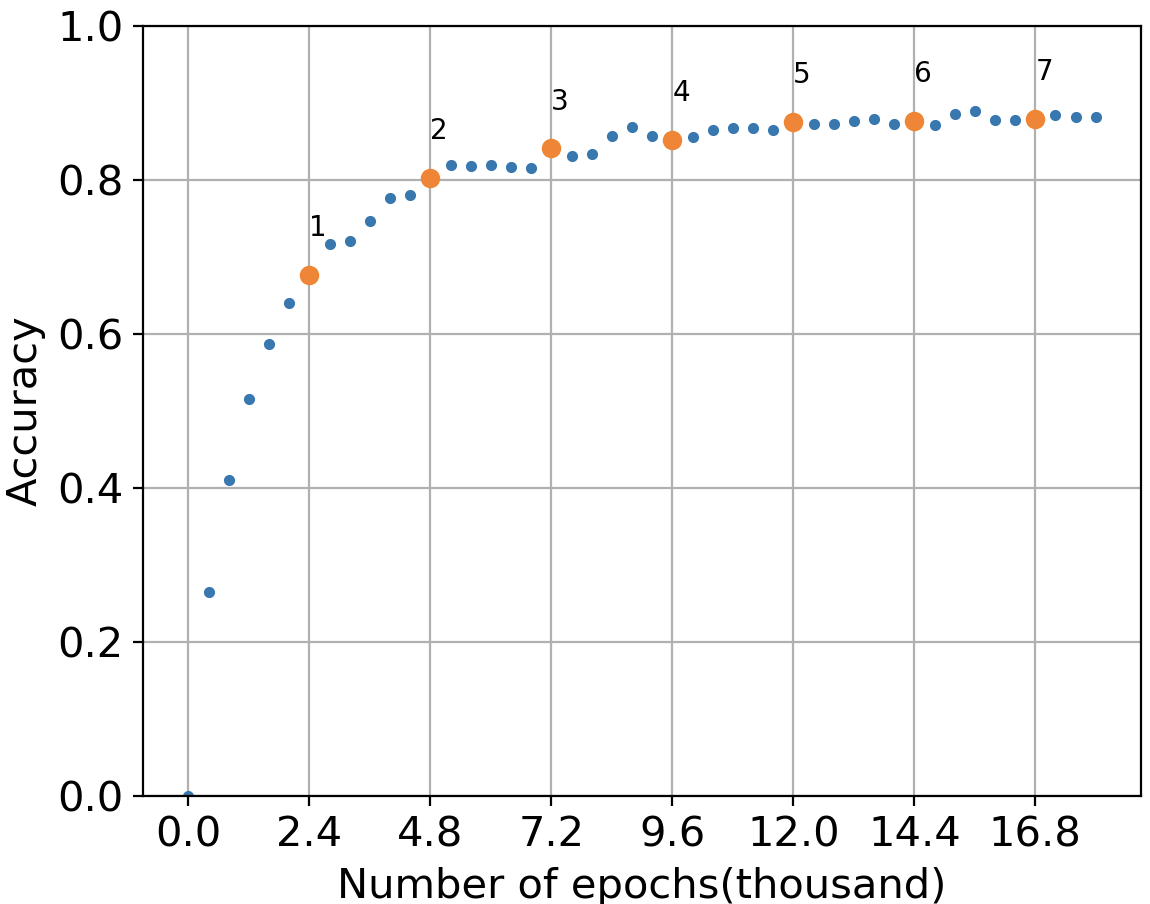}\vspace{-10pt}
    \caption{Accuracy increment across blocks and epochs}
    \label{fig:accuracy-per-block}\vspace{-10pt}
\end{figure}

We also performed a simulation to see how accuracy of a model increases with the growth of blockchain. Specifically, we measured the accuracy increment along the epochs in the DL training by measuring the accuracy of the model we achieved every 400 epochs, which takes approximately 100 seconds. In Bitcoin, this translates to 1 block per 2400 epochs, and the result is shown in Fig.  \ref{fig:accuracy-per-block}. For more complicated models, the training will span across more blocks.

%\noteTaeho{Need total epochs and total time for one training by Boyang}

%\noteTaeho{Changhao needs to explain how many epochs translate to one block}

%To see the accuracy gain of a model over a sequence of new blocks, we chose 40 checkpoints every 400 epochs and evaluated the accuracy on each of them. Fig. \ref{fig:exp2} shows our results and we can learn from the chart that the accuracy increased quickly at the first several checkpoints but slowed down later. There were also some fall backs as each step taken by the computer did not always ensure that the model must be more accurate.

\section{Conclusion and Future Work}

We presented a proof-of-concept design of a energy-recycling blockchain with our novel PoDL mechanism. Miners perform training tasks of deep learning instead of hash calculation, and they present trained DL models as their proof of deep learning. Model stealing and overfitting are prevented by our block acceptance policy with separated phases. Without majority DL training power, double spending is hard even though majority of full nodes are malicious. 

%Our blockchain design is compatible with all cryptocurrencies that are based on hash-based PoW mechanisms. 
Our proof-of-concept design has much room of improvement. %It is our future work to study more deeply about the impact of different deep learning models and dataset distributions to the properties of blockchain (\textit{e.g.,} number of blocks necessary for high-enough accuracy, probability of successful double spending based on accuracy, and more extensive evaluation with real-world datasets). Although the overhead brought by and the security of PoDL has been analyzed in previous section, 
The model requester may be generalized to multiple malicious requesters who may collude with miners.
%there can be multiple model requesters involved, some of them may be malicious and collude with each other or even with miners, they may wish to remove datasets after they receive finally trained model, \textit{etc.} 
Besides, more extensive study needs to be performed  with a realistic pattern of block submission and more DL models/datasets. It is our future work to extend this study to improve and complete the PoDL mechanism.

%for the impact of different deep learning models and dataset distributions to the properties of blockchain (\textit{e.g.,} number of blocks necessary for high-enough accuracy, probability of successful double spending based on accuracy, impact of network delay caused by model propagation). 

%our mechanism do have various limitations including how to keep training and test datasets hosting as the model requester should have plenty of dataset to propagate and store, how to make sure the training dataset is correctly and firmly broadcasted to the whole network, we also need to store the model into the block for validation purpose and thus the ratio of transactions in a block will dependent on the size of the model written in the block.

%\noteTaeho{Need to admit various limitations. 1: training/test datset hosting; 2: training dataset propagation; 3: model size; 4: }

\newpage

\bibliographystyle{splncs04}
\bibliography{main}

\begin{thebibliography}{10}
\providecommand{\url}[1]{\texttt{#1}}
\providecommand{\urlprefix}{URL }
\providecommand{\doi}[1]{https://doi.org/#1}

\bibitem{gridcoin-white-paper}
\url{https://gridcoin.us/assets/img/whitepaper.pdf}

\bibitem{hash-benchmark}
\url{https://www.febooti.com/products/filetweak/members/hash-and-crc/hash-benchmark/}

\bibitem{digiconomist}
Bitcoin energy consumption index,
  \url{https://digiconomist.net/bitcoin-energy-consumption}

\bibitem{burstcoin}
Blockchain, \url{https://www.burst-coin.org/proof-of-capacity}

\bibitem{nxt}
The blockchain application platform, \url{https://nxtplatform.org/}

\bibitem{Foldingcoin}
Foldingcoin, \url{https://foldingcoin.net/}

\bibitem{home-golem}
Golem gnt, \url{https://golem.network/}

\bibitem{nem}
Nem, \url{https://nem.io/technology/}

\bibitem{sorting-benchmark}
Comparison of internal sorting algorithms (Sep 2008),
  \url{https://attractivechaos.wordpress.com/2008/08/28/comparison-of-internal-sorting-algorithms/}

\bibitem{hashtable-benchmark}
 (Aug 2016),
  \url{https://tessil.github.io/2016/08/29/benchmark-hopscotch-map.html}

\bibitem{coppola_2018}
Coppola, F.: Bitcoin's need for electricity is its 'achilles heel' (May 2018),
  \url{https://www.forbes.com/sites/francescoppola/2018/05/30/bitcoins-need-for-electricity-is-its-achilles-heel/\#3825e5ef2fb1}

\bibitem{dwork1992pricing}
Dwork, C., Naor, M.: Pricing via processing or combatting junk mail. In: Annual
  International Cryptology Conference. pp. 139--147. Springer (1992)

\bibitem{dvf2017}
van Flymen, D.: Learn blockchains by building one.
  \url{https://github.com/dvf/blockchain} (2017)

\bibitem{gong2014compressing}
Gong, Y., Liu, L., Yang, M., Bourdev, L.: Compressing deep convolutional
  networks using vector quantization. arXiv preprint arXiv:1412.6115  (2014)

\bibitem{han2015deep}
Han, S., Mao, H., Dally, W.J.: Deep compression: Compressing deep neural
  networks with pruning, trained quantization and huffman coding. arXiv
  preprint arXiv:1510.00149  (2015)

\bibitem{hern_2018}
Hern, A.: Bitcoin's energy usage is huge – we can't afford to ignore it (Jan
  2018),
  \url{https://www.theguardian.com/technology/2018/jan/17/bitcoin-electricity-usage-huge-climate-cryptocurrency}

\bibitem{king2013primecoin}
King, S.: Primecoin: Cryptocurrency with prime number proof-of-work. July 7th
  (2013)

\bibitem{lecun2015deep}
LeCun, Y., Bengio, Y., Hinton, G.: Deep learning. nature  \textbf{521}(7553),
  ~436 (2015)

\bibitem{warden2018speech}
Warden, P.: Speech commands: A dataset for limited-vocabulary speech
  recognition. arXiv preprint arXiv:1804.03209  (2018)

\end{thebibliography}

\end{document}